**Anticathode effect on electron kinetics in electron beam generated E × B plasma**

Nirbhav Singh Chopra[1,2], Ivan Romadanov[1], and Yevgeny Raitses[1]

[1] Princeton Plasma Physics Laboratory, Princeton University, Princeton, NJ 08543, United States of America

[2] Department of Astrophysical Sciences, Princeton University, Princeton, NJ 08544, United States of Americas

**Abstract**

Electron beam (e-beam) generated plasmas with applied crossed electric and magnetic (E × B) fields are promising for low damage processing of materials with applications to microelectronics and quantum information systems. In cylindrical e-beam E × B plasmas, radial confinement of electrons and ions is achieved by an axial magnetic field and radial electric field, respectively. To control the axial confinement of electrons, such e-beam generated plasma sources may incorporate a conducting boundary known as an anticathode, which is placed on the axially opposite side of the plasma from the cathode. In this work, it is shown that varying the anticathode voltage bias can control the degree to which the anticathode collects or repels incident electrons, allowing control of warm electron (electron energies in 10-30 eV range) and beam electron population confinement. It is suggested that the effect of the anticathode bias on the formation of these distinct electron populations is also associated with the transition between weak turbulence and strong Langmuir turbulence.

I. **Introduction**

Electron beam (e-beam) plasma sources are capable of selectively generating reactive species for material surface processing while maintaining low energetic particle flux to substrates placed in the periphery of the plasma region [1–3]. Such remote low temperature plasma sources have already demonstrated their applicability for atomic scale etching [4,5] and nanomaterial processing [6].

In many e-beam plasmas with applied crossed electric and magnetic (E × B) fields, such as Penning source, Hall thruster, magnetron, and reflex arc plasmas, ions are unmagnetized. Therefore, ion transport in these plasmas is dominated by the electric field and collisions with neutrals. Previous studies on e-beam generated E × B plasmas have proposed and demonstrated the formation of an ion-confining radial electric potential well [6–10]. This potential well may be leveraged for low-damage threshold material processing applications by limiting energetic ion flux to the plasma periphery, where substrate surface modification by radicals typically takes place. However, several ion heating mechanisms may occur, which can allow ions to overcome the radial potential well and damage peripherally-placed substrates [9].

The electron cross-field transport (perpendicular to the magnetic field) is closely related to the formation of the radial potential well [8–10], and is often fluctuation driven (i.e., anomalous) in e-beam generated E × B plasmas [9,11–15], especially at lower neutral pressures (< 1 mTorr) [16]. Under fluctuation induced transport conditions, electrons scatter not only with neutral atoms but also turbulent plasma structures and oscillations, for instance plasma density and potential fluctuations [7,8,17,18]. Electron transport enhanced by fluctuations can short the ion-confining electric field and is therefore another mechanism that can introduce ions to the peripheral substrate processing region. Therefore, quantifying the anomalous electron transport is critical in developing physical understanding of different operating regimes in e-beam E × B plasmas.

Fluctuations and instabilities can also modify the electron kinetics. Weakly collisional e-beam plasmas (neutral pressure < 100 mTorr) are susceptible to beam plasma instabilities (BPIs), which can generate so-called 'warm' electrons with energies that are intermediate between the injected beam energy and the bulk electron temperature [19,20]. For sub-100 eV e-beam energies injected into a plasma with electron temperature of ~10 eV, it has been shown that a BPI induced warm electron population can have energies in the 10-30 eV range [19–21], coinciding with the typical maxima of electron impact excitation and dissociation cross sections of neutral species [22–26]. Therefore, the onset of BPIs may be important to consider for controlling the production of reactive species in the plasma.

A typical e-beam plasma source consists of an electron emitting cathode (e.g. thermionic [8], ion-induced secondary electron emitting [3,27], or rf plasma [7] cathode) mounted at one end of an anodic cylindrical vacuum chamber filled with neutral gas. The emitted electrons ionize the neutrals, producing bulk electrons and ions. An externally applied axial magnetic field radially confines the resulting e-beam and bulk plasma electrons. A tertiary electrode (often called an 'anticathode') is often placed on the axially opposite side of the plasma from the cathode, and is biased to the anode potential so that it collects beam and bulk electrons [27–30]. One modification of this conventional e-beam plasma source is the reflex arc configuration [10,30,31], in which the anticathode is biased to the cathode potential rather than the anode potential. In such a configuration the anticathode reflects incident beam and bulk electrons back into the plasma volume, thereby increasing the plasma density. However, it remains unclear whether the increase in plasma density produced by the anticathode is due to an increase in ionization rate or reduction in axial electron losses.

There have been several recent studies investigating the effect of the axial boundary opposite the cathode on the macroscopic and transport properties of similar devices. Previous work on a E × B Penning plasma generated by non-thermal electrons demonstrated that the cross field electron transport is strongly affected by whether the axial boundary opposite to the cathode is an electron-collecting conductor or an electron-repelling dielectric [7]. Furthermore, it has been shown in an e-beam generated plasma produced in a dielectric flask that electrons incident on the axial boundary can produce a significant electron induced secondary electron emission (eSEE) flux back into the plasma [28,29]. However, to our knowledge, the effect of the voltage bias applied to the anticathode on the electron energy distribution function (EEDF) in the e-beam generated E × B plasma system has not yet been experimentally quantified.

In the present work, we explore the effect of a variable voltage bias anticathode on electron kinetics in an e-beam generated E × B plasma. Here the anticathode is biased to either repel or collect incident electrons, allowing control of the axial confinement of electrons in the discharge. The radial dependence of the EEDF and resultant electron density and temperature is determined using a Langmuir probe diagnostic, while the plasma potential profile is determined using a floating emissive probe technique. Results indicate that the anticathode bias allows control of the core plasma density over nearly an order of magnitude. Additionally, the increase in plasma density for an electron-repelling bias voltage of the anticathode is mainly due to reduced axial plasma losses rather than enhanced ionization. Furthermore, a warm electron population is created in the vicinity of the plasma axis when the anticathode is electron collecting.

The paper is organized as follows: Section II discusses the experimental setup of the e-beam generated E × B plasma. Section III discusses the plasma diagnostics and measurement procedures used. Section IV details the results of the probe measurements and their analysis. Models of the dependence of electron confinement and warm electron production on the anticathode bias are proposed and compared to the experimental measurements in Section V. Conclusions are summarized in Section VI.

## II. Experimental setup

The experimental setup is depicted in Figure 1. In addition, Table I lists key geometrical and operating parameters of this setup and characteristic dimensionless parameters of the studied plasma regimes. The e-beam chamber consists of a cylindrical vacuum vessel that is pumped by a turbomolecular pump backed by a rough pump to a base pressure of ~1 $\mu$Torr. The chamber is filled with Ar gas to a pressure of 0.1-0.5 mTorr. An axial magnetic field of 50-100 G is generated by electromagnets arranged in a Helmholtz-like configuration along the chamber. A thermionic cathode consisting of an ohmically heated tungsten filament is placed on one end of the cylindrical chamber. The tungsten filament is made from a 0.4 mm diameter wire with exposed length of 1 cm [8,25]. The cathode is mounted on a floating ceramic break with a 1.75 cm inner radius, such that the cathode is electrically isolated from the grounded chamber. Therefore, a portion of the wall directly behind the cathode filament is electrically floating, with a corresponding radius $r < 1.75$ cm. This prevents electron loss to the wall behind the cathode (Figure 1), and hence the region of $r < 1.75$ cm will be called the cathode region.

The cathode is biased to a potential $V_c$ that is negative relative to the grounded chamber (anode, defined as 0V), injecting the emitted electrons into the chamber as a nonthermal electron population. Most of the applied voltage drop occurs in the cathode sheath, and thus electrons emitted from the cathode are axially accelerated by the cathode sheath into the chamber as a beam with energy $\varepsilon_b \approx e(V_{pl} - V_c)$, where $V_{pl}$ is the plasma potential. Therefore, the discharge voltage is defined as $V_d = -V_c$. A portion of the beam electrons injected into the chamber from the cathode collide with the neutral gas atoms, ionizing them and forming ions and bulk plasma electrons with lower energies than the beam electrons. An electrically isolated anticathode made from a non-magnetic stainless-steel plate with radius $R_{atc} = 4.5$ cm is installed on the end of the chamber opposite to the thermionic cathode.

The anticathode is biased to a voltage $V_{atc}$ relative to the anode potential. By varying the anticathode bias voltage, the anticathode may operate in one of two nominal modes: i) biased at the cathode potential (repeller mode) and ii) biased at the anode potential (collector mode). In the former mode, the anticathode repels plasma and beam electrons back into the plasma volume while in the latter mode, the anticathode collects nearly all incident electrons. The anticathode may also be biased to an arbitrary potential. When the anticathode voltage bias is between the cathode and anode potentials, $V_c < V_{atc} < 0$ V, only lower energy electrons with incident energies $\varepsilon < \sim e(V_{pl} - V_{atc})$ are reflected by the anticathode sheath back into the plasma.

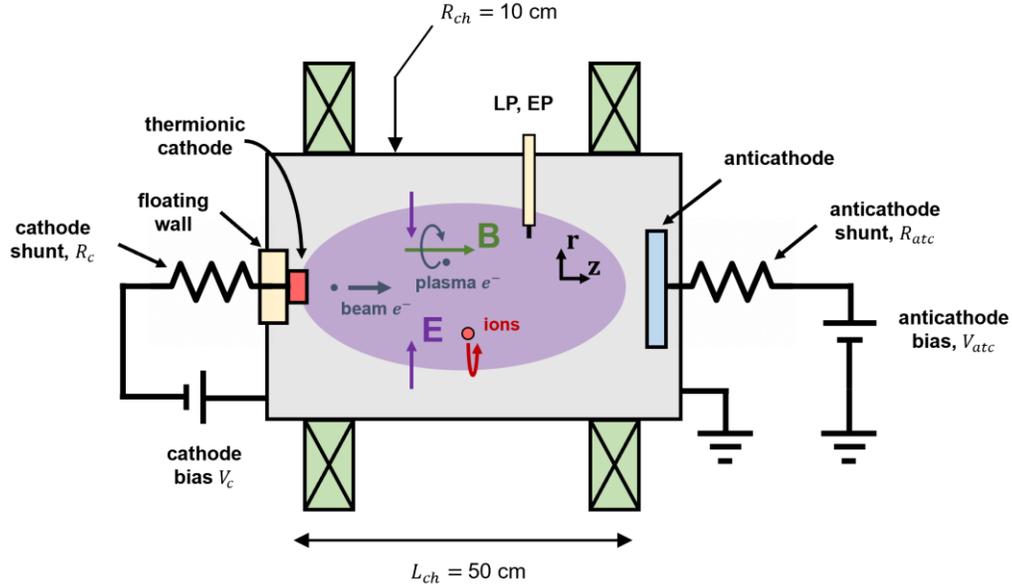

**Figure 1.** Schematic of experimental setup in $r - z$ plane.

**Table I.** Setup geometry, operating parameters, and dimensionless relationships. Symbols used in the table: $L_{ch}$ — chamber axial length, measured from cathode to anticathode; $R_{ch}$ — chamber radius; $R_{atc}$ — anticathode radius; $R_{fl}$ — radius of floating wall behind cathode; $B$ —magnetic field strength; $p$ — neutral gas pressure; $V_c$ —cathode potential, measured relative to anode (chamber) potential; $I_d$ — discharge current; $P_{heater}$ — Ohmic power supplied to heat thermionic cathode; $n_e$ — electron density [9]; $T_e$ — electron temperature [9]; $r_{L,e}, r_{L,i}$ — electron and ion Larmor radius, respectively; $R_w$ — Langmuir and emissive probe wire radius; $\nu_{en,el}, \nu_{en,inel}, \nu_{en,iz}$ — electron-neutral elastic, inelastic, and ionization collision rate for beam electrons impacting Argon, respectively (cross sections data from Ref. [32]); $\nu_{tr}$ — inverse e-beam transit time over chamber length given by $\nu_{tr} = v_b/L_{ch}$, where $v_b = (2\varepsilon_b/m_e)^{1/2}$ is the e-beam mean velocity.

| Property | Value |
|---|---|
| $L_{ch}$ | 50 cm |
| $R_{ch}$ | 10 cm |
| $R_{atc}$ | 4.5 cm |
| $R_{fl}$ | 1.75 cm |
| $B$ | 50-100 G |
| $p$ | 0.1-0.5 mTorr |
| $V_c$ | -55 V |
| $i_d$ | 50-100 mA |
| $P_{heater}$ | 42 W |
| $n_e$ | $10^9 - 10^{10}$ cm$^{-3}$ for $r = 0$ cm |
| $T_e$ | 5-12 eV for $r = 0$ cm |
| $r_{L,e}/R_{ch}$ | $5 \times 10^{-3}$ for 100 G, $T_e = 5$ eV |
| $r_{L,i}/R_{ch}$ | 0.1 for 100 G, $T_i \sim 0.03$ eV |
| $r_{L,e}/R_w$ | 10 |
| $\nu_{en,el}/\nu_{tr}$ | $10^{-1}$ |
| $\nu_{en,inel}/\nu_{tr}$ | $10^{-3}$ |
| $\nu_{en,iz}/\nu_{tr}$ | $10^{-2}$ |

### III. Diagnostics and measurement procedure

The plasma voltage-current (V-I) characteristic was characterized to determine the effect of the anticathode bias voltage on the axially and radially conducted current. Current collected by the cathode and anticathode, $I_d$ and $I_{atc}$, is determined by measuring the voltage across current shunts $R_c$ and $R_{atc}$, respectively. The current shunts are placed in series with the corresponding biasing supply, each with a resistance of 15 Ω. The voltage across each shunt is measured using a Lecroy AP031 differential probe.

To characterize how the anticathode bias voltage affects the electron kinetics and macroscopic plasma parameters and transport, a cylindrical Langmuir probe (LP) and a floating emissive probe are implemented in the described experiments. Each probe is installed at the axial midplane (25 cm from the cathode) on a movable positioner to measure radial variations of the plasma properties. Thus, the tips of both probes are oriented perpendicular to the magnetic field. Supplementary data of the plasma density fluctuations is also collected using an azimuthally oriented ion probe array, described in Appendix 1.

The LP consists of a main probe nested concentrically in a reference probe. The main probe tip is 0.1 mm diameter and 2.5 mm length. It is constructed from a tungsten wire inserted in an alumina ceramic tube. This ceramic tube is nested in a molybdenum tube which acts as a reference probe for low frequency noise suppression. The radius of the LP tip was selected to be much less than electron gyroradius, $r_{L,e}$ (Table I). This allows the measured data to be analyzed using unmagnetized probe theory [33–35].

For determining the EEDF, $n_e$, and $T_e$, biasing and data acquisition of the main probe and reference probe are controlled by a Plasma Sensors Multi-Functional Probe Analyzer (MFPA) [33]. The probe bias voltage $V_B$ is swept at a frequency of 1 kHz, and the current collected by the probe $I_{pr}$ is measured by an internal shunt of the MFPA probe analyzer. For each probe acquisition, 3 sets of 10000 current-voltage (IV) traces are acquired to determine the mean and standard error of the measurement.

Electron density $n_e$ and electron temperature $T_e$ can be deduced by integrating over the EEDF measured with the LP. The plasma potential is determined by finding the maximum of $\mathrm{d}I_{pr}(V_B)/\mathrm{d}V_B$. The EEDF $f_e$ as a function of electron energy $\varepsilon$ is then determined by the Druvesteyn method [33],

$$f_e(\varepsilon) = \frac{2m_\mathrm{e}}{e^2 A_{\mathrm{pr}}} \left(\frac{2e\varepsilon}{m_\mathrm{e}}\right)^{1/2} \frac{\mathrm{d}^2 I_{pr}}{\mathrm{d}V_B^2}(V_B = \varepsilon/e), \tag{1}$$

where $\varepsilon = e(V_{pl} - V_B)$, $m_e$ is the electron mass, and $A_{pr}$ is the probe collecting area. The zeroth and first moments of $f_e(\varepsilon)$ are then computed to determine $n_e$ and $T_e$ as

$$n_\mathrm{e} = \int_0^\infty f_e(\varepsilon)\mathrm{d}\varepsilon, \tag{2}$$

$$T_\mathrm{e} = \frac{2}{3}\frac{1}{n_\mathrm{e}}\int_0^\infty \varepsilon f_e(\varepsilon)\mathrm{d}\varepsilon, \tag{3}$$

and the ionization rate constant $K_{iz}$, is computed as

$$K_{iz} = \frac{1}{n_e}\left(\frac{2}{m_e}\right)^{1/2} \int_0^\infty \mathrm{d}\varepsilon \ \sigma_{iz}(\varepsilon)\varepsilon^{\frac{1}{2}}f_e(\varepsilon), \tag{4}$$

where $\sigma_{iz}$ is the ionization cross section for electron impact ionization of argon [36].

The floating emissive probe is used to measure the plasma potential $V_{pl}$ relative to the grounded anode. The floating probe consists of a 0.1 mm diameter tungsten wire in a 1.5 mm outer diameter double-bore alumina tube. The tungsten wire forms a 3 mm diameter loop at the head of the probe, which acts as the electron emitting surface of the emissive probe. The emissive probe is heated ohmically by an external power supply, such that the flux of thermionically emitted electrons is sufficiently greater than the critical emission current necessary for the probe floating potential to saturate. Following previous works on similar E × B plasma devices [37], the emissive probe tip is therefore assumed to float at approximately the plasma potential. The floating potential of the emissive probe relative to the anode potential is measured using a Teledyne Lecroy AP031 differential probe. The offset to the measured floating potential by the ohmic heating potential drop applied over the wire is corrected by adding half of the applied heating bias potential $V_h/2$ to the measured floating potential ($V_h = 5.9\ \mathrm{V} \pm 0.05\ \mathrm{V}$), as was performed in previous experimental works [38,39]. An additional corroborating measurement of $V_{pl}$ is performed independently using the LP, which is further discussed in Appendix 2.

## IV. Experimental results
## 1. Voltage-current (V-I) characteristics

Figure 2 shows variations of the discharge current with the discharge voltage of the plasma source for the anticathode operating in the repeller and collector modes. Results are shown for magnetic field $B = 50$ and $100$ G and neutral argon pressure $p = 0.1$ and $0.5$ mTorr. With all other parameters kept the same, the discharge current increases with increasing magnetic field and pressure. For the $p = 0.1$ mTorr case, the discharge current saturates at discharge voltages of $V_d > \sim 50$ V. For these regimes, the discharge current is limited by thermionic emission, i.e., by the temperature of the cathode filament. This emission limited regime was confirmed in a separate experiment when the discharge current was increased from 50 mA to 152 mA when the cathode heating power was increased from 38W to 45W, while all other parameters were held the same.

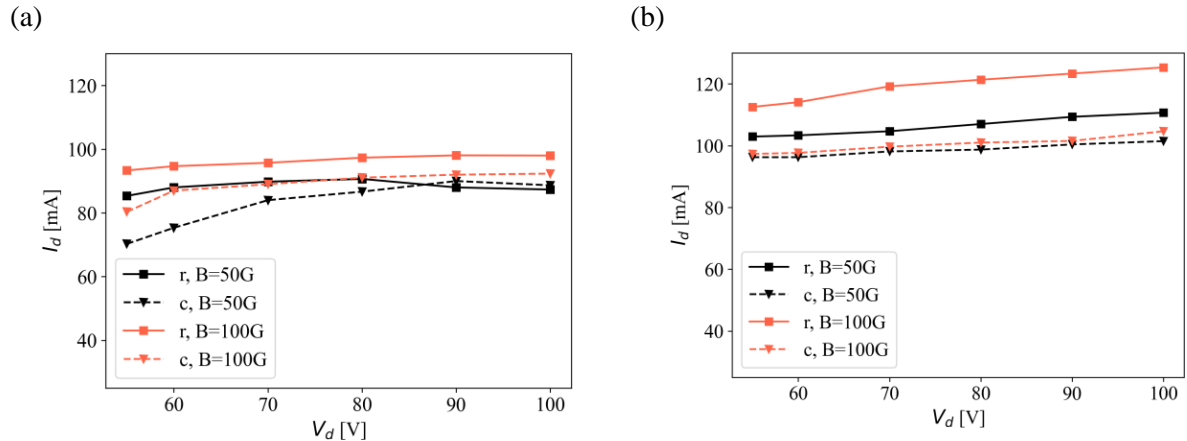

**Figure 2.** Discharge current vs. discharge voltage for repeller (r) and collector (c) modes for $p = 0.1$ mTorr (a) and $p = 0.5$ mTorr (b).

For the analysis of the V-I characteristics of the discharge (Figure 2) we introduce the ratio $x = I_{atc}/I_d$ which characterizes the fraction of the discharge current collected by the anticathode (Figure 3). Here, we assumed that the contribution of the ion current to the discharge current at the cathode is negligible. This implies that in the limit of $x \approx 1$, the flux of beam electrons emitted from the cathode and the net electron flux collected by the anticathode are equal and that there is no net current conducted across the magnetic field. From Figure 3, it is evident that for collector mode, 40-70% of the current is conducted axially towards the anticathode, while 30-60% is conducted either radially or axially to the chamber wall adjacent to the cathode. Negative values of the current ratio $x$ measured for the repeller mode implies the current collected by the anticathode is of opposite polarity compared to that of the collector mode. This is because in this mode, the anticathode current is driven by ion flux from the plasma to the anticathode, as well as by electron flux emitted from the anticathode. Since the anticathode is a cold electrode, any electron flux evolved from the anticathode surface in the repeller mode is due to ion-induced secondary electron emission (iSEE).

For the repeller mode, ions from the plasma are accelerated by the voltage drop across the anticathode sheath and strike the anticathode surface with an energy of $\varepsilon_i \approx -eV_c$. Since the voltage bias applied to the cathode considered in this work are limited to $|V_c| < 100$ V, the ion energy incident on the anticathode is $\varepsilon_i < 100$ eV. For such low energy Ar ions striking a stainless steel, the iSEE yield is much less than unity [40,41]. As a result, the iSEE evolved from the anticathode is negligible. This explains why the axial current fraction in repeller mode is observed to be $x \approx 0$.

For the collector mode with $x \approx 1$, the anticathode current is driven mainly by the electron flux consisting of beam electrons and bulk plasma electrons. In this mode, the fraction of the current conducted axially towards the anticathode increases with the magnetic field and gas pressure. Both trends are likely associated with the improved beam and plasma electron confinement in the radial direction (transverse to the magnetic field) as magnetic field strength and pressure increase. Enhanced electron confinement in the higher-pressure regime is likely due to reduced plasma density oscillations (Appendix 1), which correspond to suppressed cross-field fluctuation induced electron transport [11].

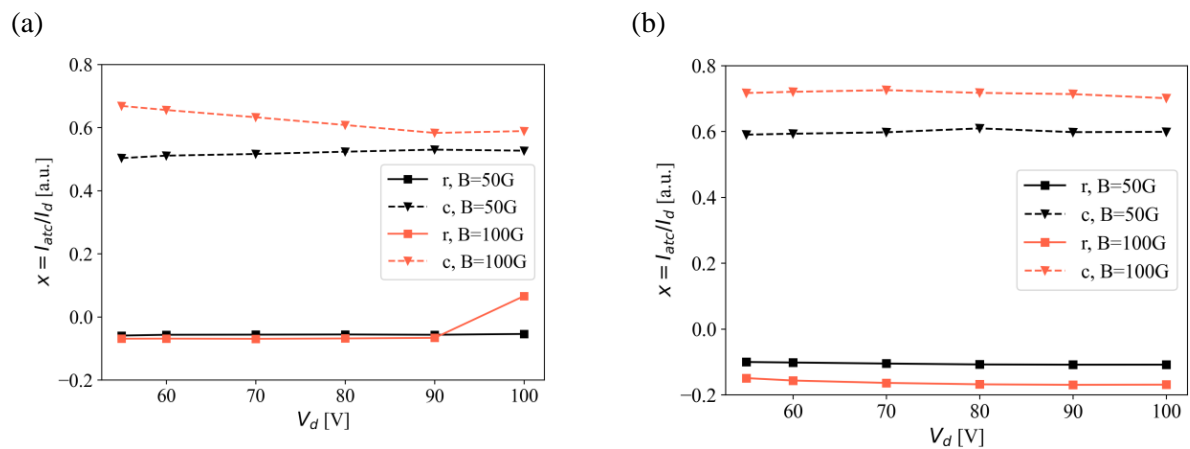

**Figure 3.** Axial current ratio $x$ vs. discharge voltage for $p = 0.1$ mTorr (a) and $p = 0.5$ mTorr (b).

Figure 4 shows the effect of the anticathode bias voltage on the axial current ratio and discharge current for the experimental parameters $p = 0.1$ mTorr, $B = 100$ G, and $V_d = 55$ V. The axial current fraction can be controlled to be nearly $x \approx 1$ in collector mode and $x \approx 0$ in repeller mode. The discharge current saturates at $I_d \approx 90$ mA as the applied voltage bias to the anticathode decreases (i.e. for a more electron repelling anticathode). The discharge current and axial current fraction both increase as the anticathode voltage bias is increased above the chamber potential, $V_{atc} > 0$ V. When the voltage bias applied to the anticathode is positive, the anticathode becomes the anode instead of the grounded chamber walls. This effectively increases the voltage drop in the cathode sheath. Therefore, the injected e-beam has an increased energy and hence is better at ionization, which leads to an increased discharge current.

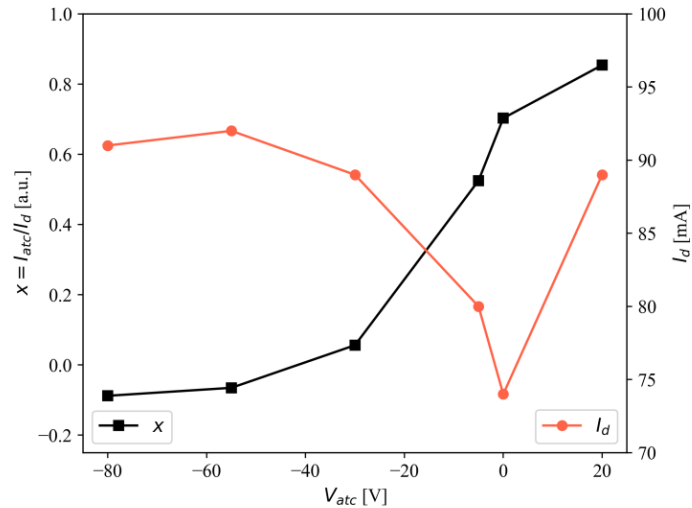

**Figure 4.** Axial current ratio $x$ and discharge current $I_d$ vs. anticathode potential $V_{atc}$ for $p = 0.1$ mTorr, $B = 100\ G$, $V_d = 55$ V.

## 2. Plasma properties

Figure 5 shows EEDFs measured at the center of the plasma source, $r = 0$ cm, in the plasma region with the electron beam, for experimental parameters $p = 0.1$ and 0.5 mTorr, $B = 100$ G, $V_c = -55$ V. For both pressure regimes, the area under the EEDF curve is maximal in the repeller mode, indicative of the overall increase in plasma density in repeller mode. For $p = 0.1$ mTorr, an electron beam component can be observable as bump-on-tail centered at the energy of $\varepsilon \approx 55$ eV (Figure 5a). This bump corresponds to the applied discharge voltage which corresponds to the maximum energy acquired by electrons emitted from the cathode and accelerated in the cathode sheath (e-beam energy). As the anticathode bias voltage increases to the anode potential (0 V), the population of bulk and energetic electrons reduces, as indicated by the reduction of area under the EEDF.

Note that for the collector mode at $p = 0.1$ mTorr, a warm electron population in the energy range of $10 - 30$ eV is formed, as indicated by the plateau in the normalized EEDF (Figure 5b). At $p = 0.5$ mTorr, the electron beam population in the EEDF is suppressed, due to the higher collision frequency of beam electrons scattering with neutrals. This effect is further quantified in Section V.1.

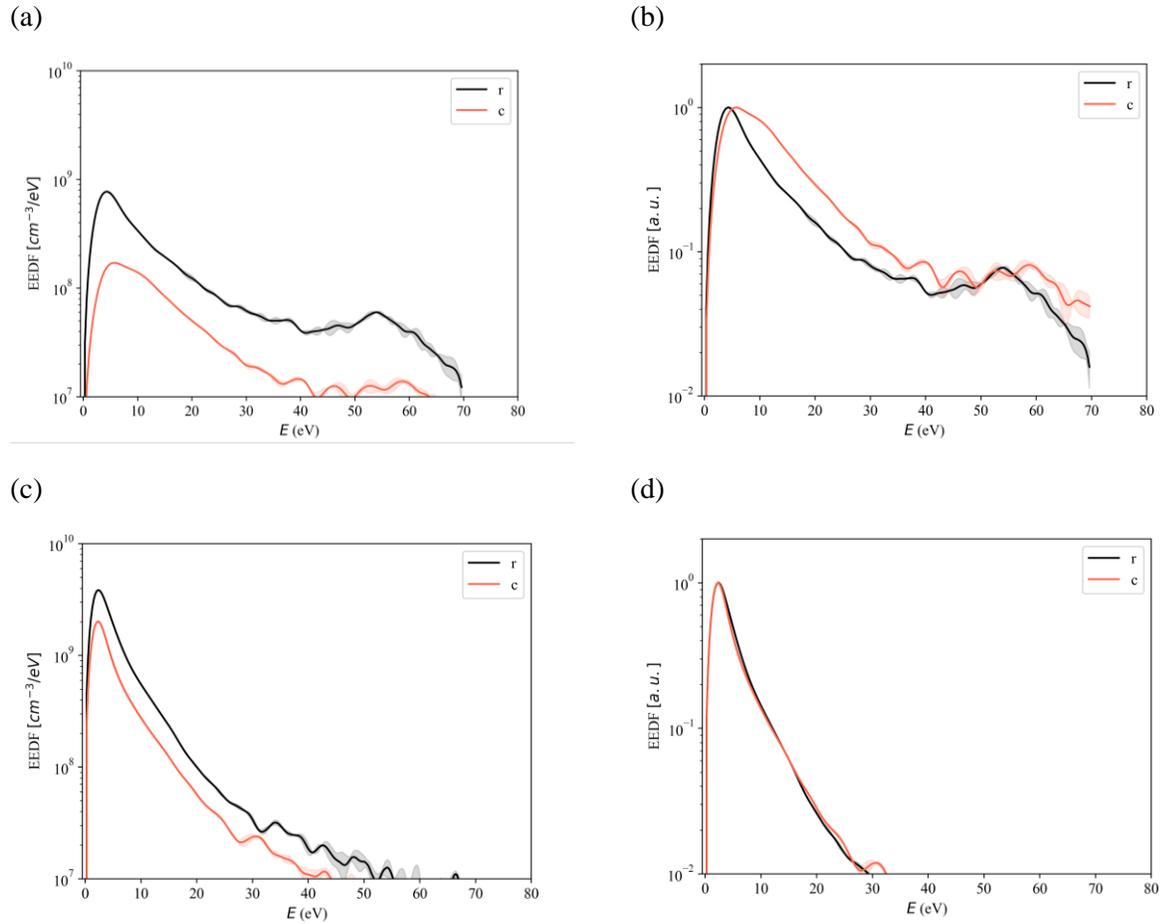

**Figure 5.** EEDF determined by LP in repeller (r) and collector (c) modes at $B = 100$ G, $V_c = 55$ V, for (a) $p =0.1$ mTorr in absolute units, (b) $p =0.1$ mTorr, normalized to EEDF maximum, (c) $p =0.5$ mTorr in absolute units, (d) $p =0.5$ mTorr, normalized to EEDF maximum.

Figure 6 shows spatial variations of the EEDF for both the repeller and collector modes for operating parameters $B = 100$ G, $p = 0.1$ mTorr, $V_c = -55$ V. For the repeller mode, an e-beam component is only observable at $r = 0$ cm. In both these modes, the beam electron population decays rapidly away from the discharge center, and the EEDF becomes approximately Maxwellian for radii $r \geq 1$ cm.

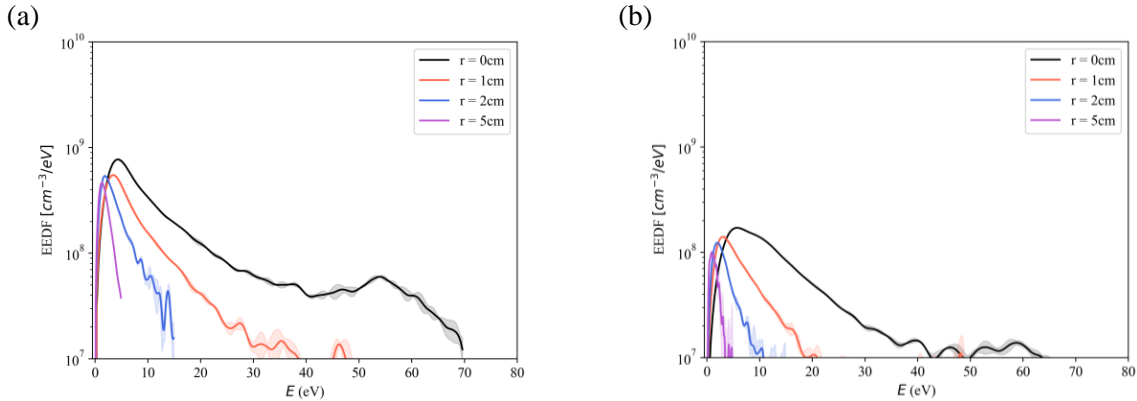

**Figure 6.** EEDFs determined by LP at varying radii at $B = 100$ G, $V_c = -55$ V, for $p = 0.1$ mTorr, for repeller mode $V_{atc} = -55$ V (a) and collector mode $V_{atc} = 0$ V (b).

Macroscopic plasma properties $n_e$ and $T_e$ deduced from the measured EEDFs are shown in Figure 7 and Figure 8. The anticathode bias voltage has a strong effect on the bulk plasma density (Figure 7), but almost no effect on the electron temperature (Figure 8). The measurement of plasma density throughout the whole region $r = 0 - 5$ cm indicates that the bulk electrons produced in the beam region ($r = 0$ cm) can diffuse across the magnetic field to populate plasma periphery at $r > 1$ cm. In the repeller mode, the electron density is larger than in the collector mode for the entire measured region, $r = 0 - 5$ cm. This enhancement of the electron density is more pronounced at $p = 0.1$ mTorr. At $p = 0.5$ mTorr, the core electron temperature reduces by a factor of ~2, likely due to increased electron-neutral collisions.

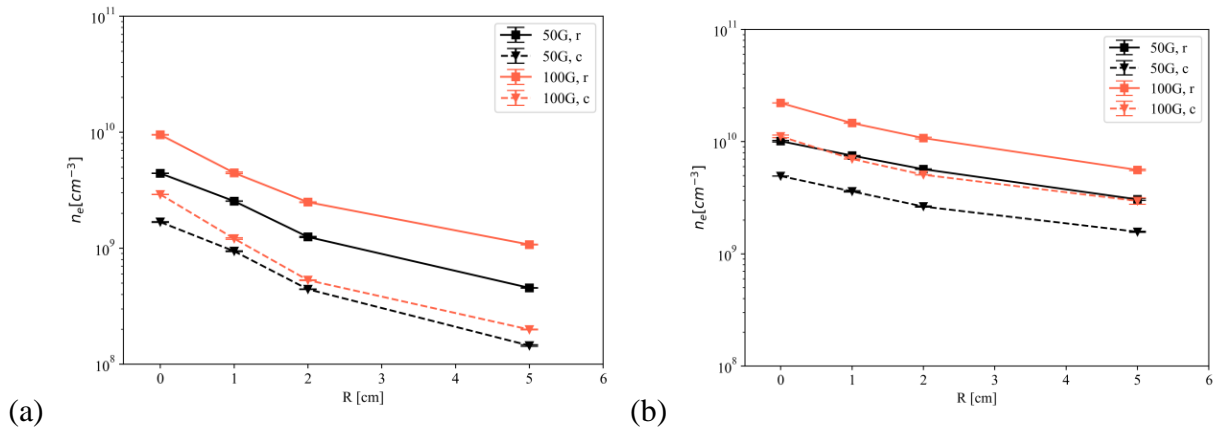

**Figure 7.** Electron density determined by LP in repeller (r) and collector (c) modes at $B = 50$, 100 G for $p = 0.1$ mTorr (a), $p = 0.5$ mTorr (b).

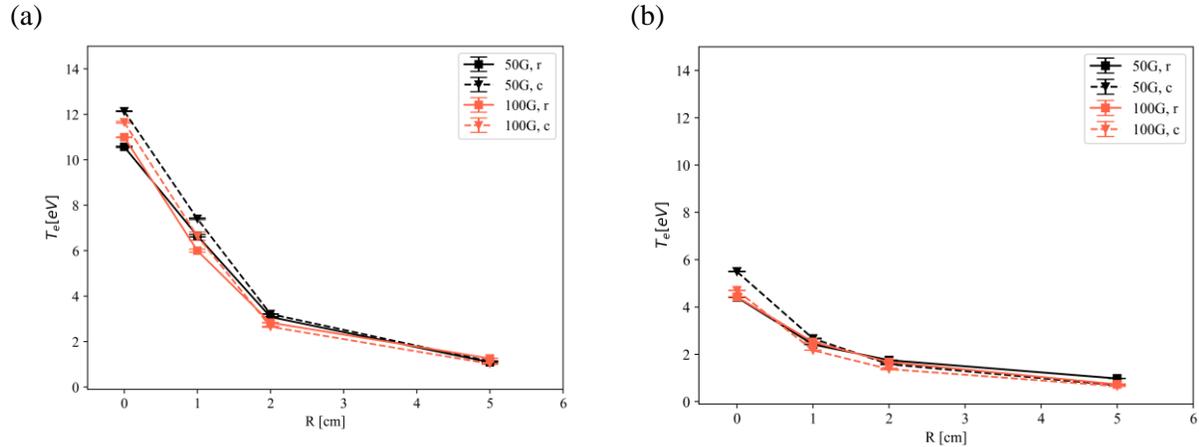

**Figure 8.** Electron temperature determined by LP in repeller (r) and collector (c) modes at $B = 50, 100$ G for $p = 0.1$ mTorr (a), $p = 0.5$ mTorr (b).

The plasma potential radial profiles $V_{pl}(r)$ are shown in Figure 9. These measurements show that the radial component of the electric field $(E_r = -\partial V_{pl}/\partial r)$ is directed radially inward. The radial electric field is observed to be weaker in the repeller mode than in the collector mode. In addition, the plasma potential in the repeller mode is lower (i.e. closer to the cathode potential) than in the collector mode. This result is consistent with the observation that the repeller mode has a better axial electron confinement than that in the collector mode. The trend of decreasing plasma potential as the magnetic field increases may occur due to enhanced radial electron confinement at larger magnetic fields. In the peripheral plasma region $r \geq 2$ cm, the plasma potential also reduces with the pressure. This is likely due to reduced radial electron transport at higher pressures, which is further discussed in Appendix 1. It is also of note that the plasma potential is positive with respect to the anode, in contrast to previous works on similar E × B plasmas observing a negative plasma potential [7,8]. This is further discussed in Appendix 2.

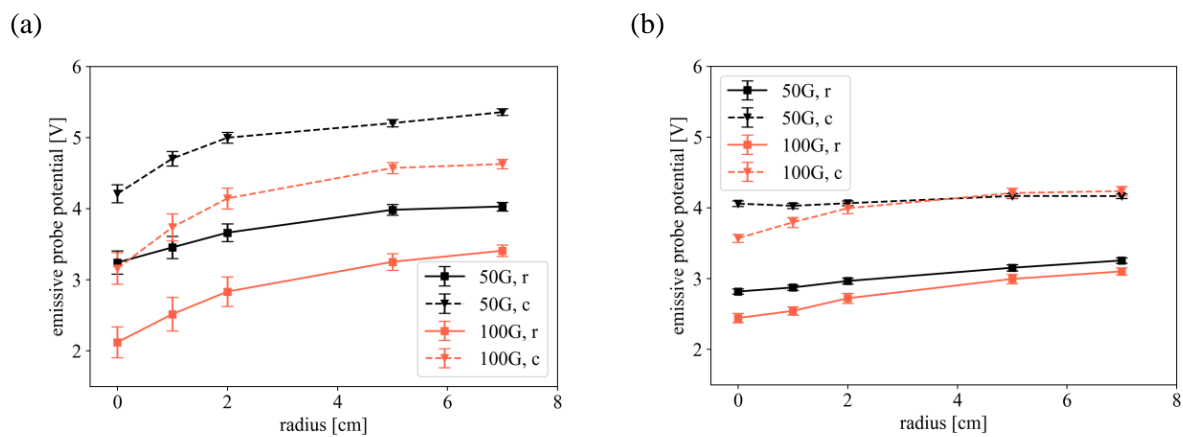

**Figure 9.** Plasma potential determined by EP at $B = 50, 100$ G for $p = 0.1$ mTorr (a) and $p = 0.5$ mTorr (b).

Figure 10 shows the radial profile of the ionization rate constant deduced from the measured EEDFs. Most of the electron impact ionization occurs for $r < 1.5$ cm, with ionization rate rapidly decaying below the detection limit of the probe diagnostic at larger radii.

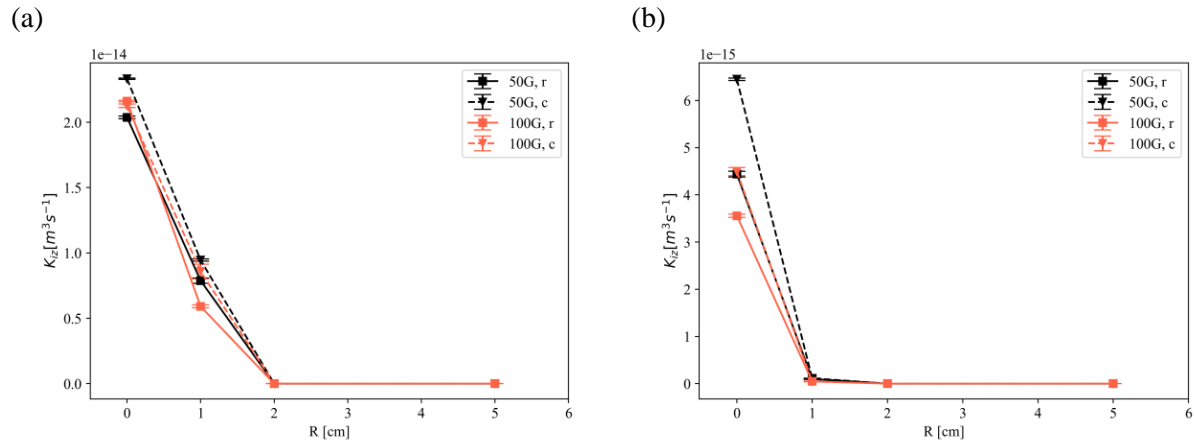

**Figure 10.** Radial profiles of experimentally determined ionization rate constants $K_{iz}$ for $p = 0.1$ mTorr (a) and $p = 0.5$ mTorr (b).

## V. Discussions

### 1. Anticathode bias effect on electron density and anomalous transport

In this section, we derive a model of the bulk electron density dependence on the applied voltage bias to the anticathode. Figure 11 shows the anticathode bias effect on the ionization rate constant at $r = 0$ cm, as determined by Eq. (4). The ionization shows negligible variation with anticathode bias. This is an important result that indicates that the EEDF produced by the electron repelling anticathode is not better at ionizing than the EEDF in collector mode. However, as will be shown, the electron repelling anticathode does have fewer axial losses, leading to an enhancement of the electron density.

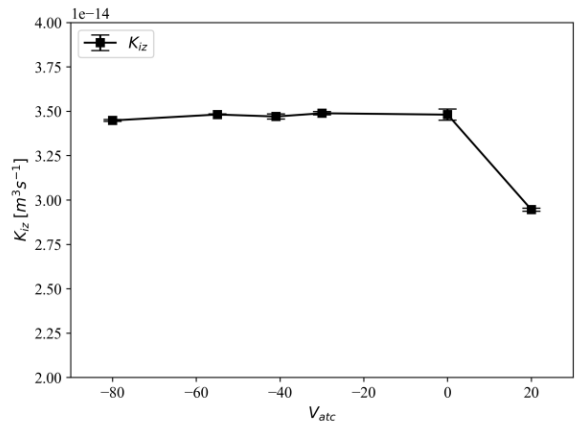

**Figure 11.** Volumetric ionization rate constant $K_{iz}$ as a function of anticathode bias potential $V_{atc}$ at $r = 0$ cm. Repeller and collector modes occur at $V_{atc} = -55$ V and 0 V respectively.

In a steady state, the electron continuity equation is given by

$$\nabla \cdot \mathbf{\Gamma}_e = R_{iz}, \tag{5}$$

where $\mathbf{\Gamma}_e$ is the electron flux and $R_{iz} = n_e n_g K_{iz}$ is the volumetric ionization rate. A diagram of the axial and radial electron fluxes in the discharge is shown in Figure 12.

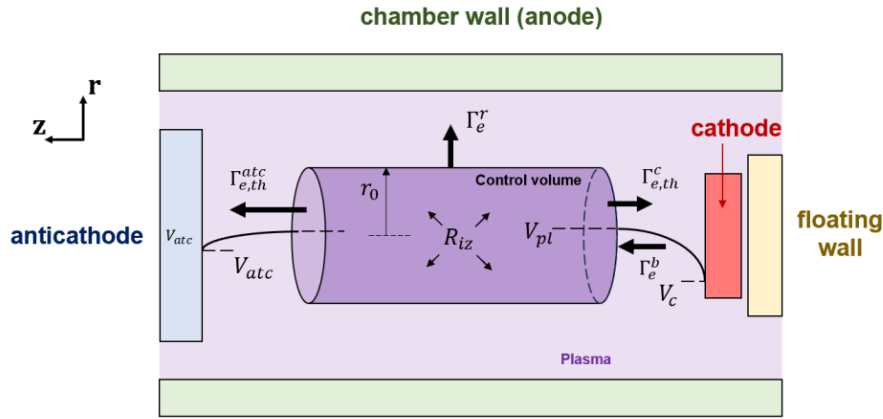

**Figure 12.** Diagram of axial and radial electron fluxes out of a cylindrical control volume of plasma, as well as volumetric ionization producing electrons in the control volume.

The electron flux leaving a control volume of plasma is balanced by beam electrons entering the volume and the production of new electrons by ionization. Consider a cylindrical control volume of radius $r_0 = 1$ cm and length $L_{ch}$. The total electron beam flux $\Gamma_e^b$ injected into a cylindrical endcap of area $A_{cap} = \pi r_0^2$ can be approximated by the discharge current $\Gamma_e^b = \frac{I_d}{eA_{cap}}$ (Figure 4). Here we assume that the contribution of the ion current to the discharge current is negligible. This assumption is justified by considering that the ion current $I_i$ collected by the cathode is equal to the Bohm current $I_i = eA_{cap}n_e(k_B T_e/m_i)^{\frac{1}{2}} \approx 3$ mA, using measured values of the plasma density and electron temperature of $n_e = 10^{16}$ m$^{-3}$ and $T_e = 11$ eV (Figure 7 and Figure 8) [42]. Evidently the ion current is much less than the discharge current ($I_d \geq 75$ mA, Figure 4), and hence can be neglected.

The axial electron flux leaving the cylindrical endcaps of the control volume consists of the one-way thermal electron flux to the cathode or floating wall and anticathode [43]. Here, we assume that the floating wall behind the cathode is charged to the same potential as the cathode due to the large electron mobility along the magnetic field. Additionally, we assume that the one-way particle flux of electrons is approximately determined by the thermal flux of an approximately Maxwellian EEDF,

$$\Gamma_{e,\text{eff.}}^M = \frac{1}{4} n_e v_{the}, \tag{6}$$

with the electron thermal velocity given by $v_{the} = (8k_bT_e/\pi m_e)^{1/2}$. An important caveat for this assumption is that in the $p = 0.1$ mTorr case, the EEDFs at $r = 0$ cm have a significant non-Maxwellian component (Figure 6a). Despite this, the one-way flux given by Eq. (6) using the effective electron density and temperatures determined by Eqs. (2) and (3) is still a reasonable approximation for the actual one-way flux of electrons (Appendix 3). Finally, we assume that the short-circuit effect at the anticathode and cathode is negligible (Appendix 3) [44]. The portion of thermal electron flux reaching the cathode and anticathode, denoted $\Gamma^c_{e,th}$ and $\Gamma^{atc}_{e,th}$ respectively, are given by [42]

$$\Gamma^c_{e,th} = \Gamma^M_{e,\text{eff.}} \exp\left(-\frac{(V_{pl} - V_c)}{kT_e}\right), \tag{7}$$

$$\Gamma^{atc}_{e,th} = \Gamma^M_{e,\text{eff.}} \min\left[\exp\left(-\frac{(V_{pl} - V_{atc})}{kT_e}\right), 1\right]. \tag{8}$$

Eqs. (7) and (8) indicate that electrons reaching either the cathode or anticathode electrodes are repelled by the sheath at the corresponding electrode. The sheath voltage drop at either the cathode or anticathode is equal to the difference between the plasma potential $V_{pl}$ and the corresponding electrode potential. Because the cathode sheath drop is much larger compared to the electron temperature, most of the thermal electron flux to the cathode is reflected back into the plasma, and $\Gamma^c_{e,th} \approx 0$. However, for the collector mode, the thermal flux to the anticathode may be significant. In such a case, the one-way electron flux that strikes the anticathode may generate a significant flux of secondary electrons back into the control volume $\delta\Gamma^{atc}_{e,th}$, where $\delta$ is the eSEE yield for stainless steel. For low energy incident electrons, the eSEE coefficient depends significantly on the conditioning of the surface that electrons are incident on, and can range from $\delta \approx 0 - 1$ [45–47]. SEE may also be generated by ion and metastable Ar collisions with the anticathode and cathode [41,48,49]. However, the electron fluxes due to ion-induced and Ar metastable induced SEE are negligible (Appendix 3).

For the region $r \geq 1$ cm, the EEDF is nearly Maxwellian (Figure 6), and the radial electron flux leaving the control volume $\Gamma^r_e$ perpendicular to the magnetic field in this region can be described by an electron fluid under the drift-diffusion approximation,

$$\Gamma^r_e(\alpha) = -n_e\left(\mu_{e\perp}E_r + \frac{D_{e\perp}}{L_n}\right), \tag{9}$$

where $\mu_{e\perp}$ and $D_{e\perp} = \frac{kT_e}{e}\mu_{e\perp}$ are the electron cross-field mobility and diffusion coefficients respectively, and the radial gradient length scale is given by $L_n^{-1} = (\partial n_e/\partial r)/n_e$.

To evaluate the contribution of anomalous transport to the cross field electron mobility and diffusion fluxes, an effective electron collision frequency accounting for electron-neutral collisions and fluctuation-based scattering of electrons is expressed as [11–15]

$$\nu_{at} = \nu_{en} + \alpha\omega_{ce}, \tag{10}$$

where $\nu_{en} = n_g \sigma_{en} v_{the}$ is the electron-neutral collision frequency, $\omega_{ce} = eB/m_e$ is the electron cyclotron frequency, and $\sigma_{en}$ is the effective momentum transfer cross-section for electron collisions with Ar atoms for mean energy of bulk electrons equal to $\frac{3}{2}kT_e$.

In Eq. (10), the parameter $\alpha$ is a semi-empirical parameter characterizing the relative significance of classical to anomalous collisions, which is usually determined from experiments (e.g., $\alpha = 1/16$ corresponds to so-called Bohm diffusion [11,12,14], and $\alpha = 0$ corresponds to classical collisional transport characterized by electron-neutral collisions). Previous works have also demonstrated that in $E \times B$ devices such as Hall thrusters, the anomalous parameter can reach values much smaller than the well-known semiempirical reference value of $\alpha = 1/16$ given by Bohm et al (Ref. [50]) [15]. In the e-beam plasma considered here, the electrons are magnetized, $\frac{r_{Le}}{R_{ch}} \gg 1$ and $\frac{\omega_{ce}}{\nu_{en}} \gg 1$ where $r_{Le}$ is the electron gyro-radius. Under such conditions, the parameter $\alpha$ represents the inverse of an effective electron Hall parameter. The expression for the semiempirical electron cross-field mobility is then given by

$$\mu_{e\perp}(\alpha) = \left(\frac{e}{m_e \nu_{at}(\alpha)}\right) \frac{1}{1 + \omega_{ce}^2/\nu_{at}(\alpha)^2}. \tag{11}$$

Integrating Eq. (5) over a cylindrical control volume of radius $r_0$ and length $L_{ch}$, we arrive to

$$A_{cap}\left[\left(\Gamma_{e,th}^c + \Gamma_{e,th}^{atc}\right) - \Gamma_e^b\right] + A_{sh}\Gamma_e^r = V_{cyl}R_{iz}, \tag{12}$$

where $A_{sh} = 2\pi r_0 L_{ch}$ and $V_{cyl} = \pi r_0^2 L_{ch}$ are the shell area and volume of the cylindrical control volume, respectively. Then, the bulk electron density in the control volume is analytically determined from Eq. (12) as

$$n_e = \frac{\left(V_{cyl}R_{iz} + +\Gamma_e^b A_{cap}\right)}{A_{cap}\frac{v_{th,e}}{4}\left\{\exp\left(-\frac{(V_{pl} - V_c)}{kT_e}\right) + \min\left[\exp\left(-\frac{(V_{pl} - V_{atc})}{kT_e}\right), 1\right]\right\} - \left[\mu_{e\perp}(\alpha)E_r + \frac{D_{e\perp}(\alpha)}{L_n}\right]A_{sh}}. \tag{13}$$

All parameters in Eq. (13) are experimentally determined at $r_0 = 1$ cm for repeller and collector modes ($V_{atc} = -55V$ and $0V$ respectively) except for the anomalous parameter $\alpha$ and the eSEE coefficient $\delta$. For $V_{atc} \neq -55V$ or $0V$, the electron radial drift and diffusion velocities, $\mu_{e\perp}(\alpha)E_r$ and $D_{e\perp}(\alpha)/L_n$ respectively, are linearly interpolated using the experimentally determined values at $V_{atc} = -55V$ and $0$.

The dependence of the experimentally determined bulk electron density on the anticathode potential is shown in Figure 13 for $p = 0.1$ and $0.5$ mTorr and $B = 100$ G. Also shown is the modeled bulk electron density as determined by Eq. (13) for varying values of $\alpha$. For anticathode potentials below the anode potential, $V_{atc} < 0$ V, the thermal flux of electrons incident on the anticathode is mostly reflected by the anticathode sheath, leading to an enhancement in plasma density. In this electron repelling regime of the anticathode in the $p = 0.1$ mTorr case, a good agreement between the modelled and experimentally determined electron densities is found for anomalous electron cross-field transport with $\alpha = 1/325$. In this case $\alpha\omega_{ce} \approx 5$ MHz while $\nu_{en} \approx 700$ kHz, so that $\alpha\omega_{ce} \gg \nu_{en}$. This suggests a significant enhancement in the electron cross field transport occurs due to anomalous scattering of electrons.

The experimentally determined plasma density is minimum in the collector mode (i.e. $V_{atc} = 0V$), which is consistent with the minimization of the discharge current in the collector mode (Figure 4). In this mode the anticathode collects nearly all incident electrons, leading to a depletion of plasma density. Agreement between the modelled and experimentally determined electron density in collector mode is again found for the case of $\alpha = 1/325$ in the $p = 0.1$ mTorr case. Furthermore, for a positively biased anticathode with $V_{atc} = +20V$, better agreement between modeled and experimentally determined plasma density is found for $\alpha \approx 0$, i.e., classical cross field electron transport. This trend indicates that as the anticathode becomes more electron collecting, the anomalous cross-field radial electron transport decreases.

Notably, for $p = 0.1$ mTorr the electron-neutral collisional mean free path is $\lambda_{en} \approx 150$ cm $\gg L_{ch}$. Therefore, the effect of beam electrons scattering with neutrals is insignificant at $p = 0.1$ mTorr. However, the model overpredicts the plasma density by roughly one order of magnitude in the $p = 0.5$ mTorr case, even for the anomalous electron transport case of $\alpha = 1/325$. In such a case, the electron-argon collisional mean free path decreases to $\lambda_{en} = 1/n_g\sigma_{en} \approx 30$ cm $< L_{ch}$. Under such conditions the injected beam electrons may undergo significant scattering while transiting from the cathode to the probe measurement location at the midplane of the plasma, which would manifest as a reduction in the effective injected beam flux $\Gamma_e^b$ in Eq. (13). A larger value of $\alpha > 1/325$ and hence more anomalous cross field electron transport would allow the model to match the experimentally determined plasma density in this 0.5 mTorr case. However, separate measurements indicate that the 0.5 mTorr case has reduced plasma density fluctuations, indicating that instead $\alpha < 1/325$ (Appendix 1, Figure 15). Therefore, future works should be conducted to quantify the effect of the electron beam scattering at higher pressure conditions where electron-neutral scattering becomes significant.

We also note that eSEE from the anticathode may introduce a significant electron flux into the plasma. Indeed, an eSEE yield of $\delta = 0.3$ corresponds to eSEE coefficient for stainless steel extrapolated from Ref. [46] to the bulk electron temperature of 11 eV at the plasma center (Figure 8). Under such conditions, eSEE may lead to an enhancement in the plasma density predicted by the model, necessitating a larger value of $\alpha > 1/325$ for model and experimental agreement in collector mode. Thus, in principle, the eSEE from the anticathode can significantly enhance the effective steady state axial confinement of electrons in the plasma column for the collector mode. However, the accurate determination of the eSEE yield for the stainless steel under realistic plasma conditions used in these experiments is beyond the scope of this paper and therefore, this suggestion is based exclusively on the above-mentioned references.

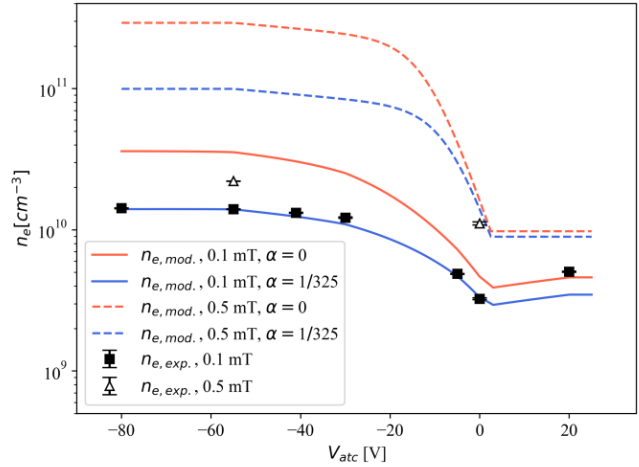

**Figure 13.** Experimentally determined electron density $n_e$ dependence on anticathode potential $V_{atc}$. Experimental parameters are $p = 0.1$ and $0.5$ mTorr, $B = 100$ G, and $V_c = -55$ V. Also shown is the modeled electron density dependence on the anticathode potential for several values of the anomalous parameter $\alpha$, as given by Eq. (13) and assuming zero eSEE yield ($\delta = 0$). Repeller and collector modes occur at $V_{atc} = -55$ V and 0 V respectively.

In summary, the model of steady state density predicted by Eq. (13) suggests when the anticathode is electron repelling, there is enhanced anomalous cross-field (radial) transport due to scattering of electrons on plasma fluctuations. This prediction is supported by separate measurements of plasma density fluctuations which indicate that the repeller mode is characterized by stronger amplitude plasma density oscillations than collector mode (Appendix 1, Figure 15a). The enhanced anomalous transport in the repeller mode may explain why the radial electric field is suppressed in the repeller mode when compared to the collector mode (Figure 9). Enhanced radial anomalous electron flux would effectively short the potential well by redistributing negative charge out of the center of the well.

2. **Remarks on beam plasma instability onset and production of warm electron population**

In Section V.1 we demonstrated the somewhat surprising result that the collector mode has a similar ionization rate constant as the repeller mode, despite the anticathode collecting nearly all incident electrons in collector mode. In this section, we will estimate the role of the anticathode bias voltage on the onset of beam plasma instabilities, which in turn may lead to the redistribution of the beam and bulk electrons to form an enhanced warm electron population that is responsible for ionization in collector mode.

The normalized EEDFs at the center of the discharge (Figure 5b) measured in the repeller mode and the collector mode suggest that for the repeller mode, there is a distinct beam electron population with $\varepsilon_b = 55$ eV, which is spatially localized to the region $r <1$ cm. This is not the case for the collector mode, where instead a warm electron population in the energy range of 10-30 eV is formed. Following Ref. [19] we hypothesize that this transition from EEDF with a bump on tail in repeller mode to an EEDF with a warm electron population may be the result of the transition between different beam-plasma instability regimes.

It is known [51] that the injected e-beam can excite electrostatic Langmuir waves in the discharge, with initial wavenumber $k$ given by the dispersion relation for Langmuir waves,

$$k = \frac{\omega_{pe}}{v_b}. \tag{14}$$

Here, $\omega_{pe} = \sqrt{e^2 n_e / \varepsilon_0 m_e}$ is the bulk electron plasma frequency. As shown in previous works [19] and as will be calculated and shown later, in the low-pressure regimes considered in this work $p = 0.1 - 0.5$ mTorr, the electron-neutral collision rate is insignificant to damp the Langmuir waves. Simulations performed in Refs. [19,21] suggest that an e-beam propagating in the so-called weak turbulent (WT) regime transfers energy to Langmuir oscillations coupled with ion acoustic waves. These oscillations contribute to a broad redistribution of electron energy in the EEDF (i.e. yielding a production of warm electrons) [19,21]. On the other hand, plasma with an e-beam propagating under so-called strong Langmuir turbulence (SLT) conditions can undergo Langmuir collapse, which preserves the electron beam via production of a nonlinear (ponderomotive) electric field. The Langmuir collapse of the plasma manifests as a bump on tail in the EEDF [19,52,53].

The beam-plasma density and energy ratios, $\frac{n_b}{n_e}$ and $\frac{\varepsilon_b}{T_e}$ respectively, characterize the turbulent regime of the beam-plasma system by quantifying the beam intensity relative to the plasma, where $n_b$ is the electron beam density. Following Ref. [19] and the associated calculation of different beam plasma instability regimes in Appendix 4, we define the regions of parameter space dominated by WT and SLT for the e-beam plasma system studied here. The calculated WT and SLT regimes in $\frac{\varepsilon_b}{T_e}$ and $\frac{n_b}{n_e}$ parameter space determined in Ref. [19] are shown in Figure 14. The results of calculations are plotted along with experimental points deduced from the measured EEDFs in this work. The experimentally determined value of the bulk plasma density $n_e$ is determined by Eq. (2), and the beam density $n_b$ is determined by

$$n_b = \int_{\varepsilon_b - \Delta \varepsilon_b}^{\varepsilon_b + \Delta \varepsilon_b} f_e(\varepsilon) d\varepsilon, \tag{15}$$

where the beam energy was taken to be $\varepsilon_b \approx -eV_c = 55$ eV and the spread in beam energy was chosen to be $\Delta \varepsilon_b = 6$ eV, based on the approximate full width half maximum of the bump on tail feature observed in repeller mode (Figure 5b).

In the low-pressure case of $p = 0.1$ mTorr, the system is close to the boundary between WT and SLT regimes (Figure 14). The applied cathode potential is held fixed in all experimental cases, and hence the beam energy is approximately constant. Furthermore, the probe measurements in the present work indicate that, for a fixed neutral pressure, the bulk electron temperature at the discharge center does not vary by more than 20% (Figure 8a and b). Therefore, the beam-plasma energy ratio $\varepsilon_b / T_e$ does not vary significantly between collector and repeller mode for a fixed pressure. However, as shown on Figure 14, the beam-plasma density ratio $n_b/n_e$ does vary significantly, and therefore a transition between WT and SLT regimes may be achieved by varying the beam-plasma density ratio.

In the case of collector mode at $p = 0.1$ mTorr and $B = 100$ G, the normalized EEDF is more populated in the energy range of 10-30 eV than in repeller mode (Figure 5b). As a result, the density ratio $n_b/n_e$ is smaller in the collector mode. Collector mode evidently operates in the WT regime, which is consistent with the broad energization observed in measured EEDF (Figure 5b). Conversely, by operating in repeller mode at the same pressure and magnetic field, the beam-plasma density ratio increases due to the pronounced presence of a bump-on-tail at roughly the beam energy. Evidently this beam-plasma density ratio is sufficiently large to push the system into the SLT regime. Under SLT conditions, the phenomenon of Langmuir collapse should occur, preserving the beam population and producing a bump on tail in the EEDF [19,52,54,55]. At the higher-pressure regime of $p = 0.5$mTorr, the beam-plasma system is further from the WT/SLT boundary in parameter space, which coincides with the relatively broad EEDF observed in Figure 5d.

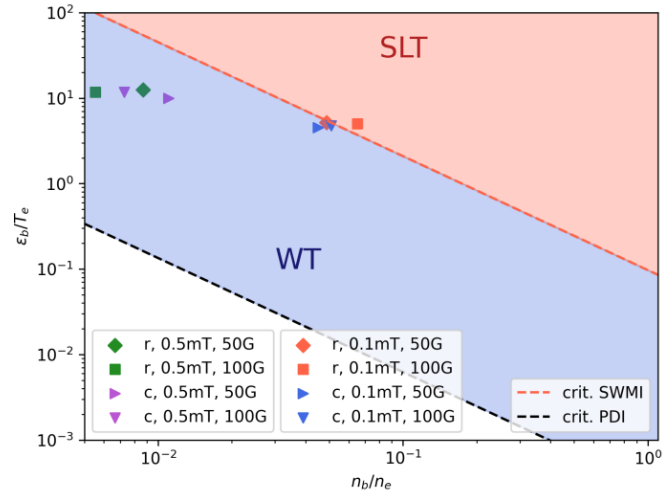

**Figure 14.** Diagram of instability regions in normalized beam energy $\varepsilon_b/T_e$ and density $n_b/n_e$ parameter space. Following Ref. [19], shaded regions shown are characterized by weak turbulence (denoted 'WT') and strong Langmuir turbulence (denoted 'SLT') regimes. Dashed lines indicate the stability boundaries for the parametric decay and standing wave modulational instabilities that define the WT and SLT regimes given by Eqs. (23) and (24) respectively (Appendix 4, Ref. [19]). Also plotted are points from experimental determination from this work of EEDFs for repeller ('r', $V_{atc} = -55\ V$) and collector ('c', $V_{atc} = 0\ V$) modes. The measured results are for $V_c = -55$ V and $I_d$ =64-70 mA. Results were obtained at two different magnetic fields, $B$ =50 G and 100 G and two different neutral pressures, $p$ =0.1 and 0.5 mTorr.

**VI. Conclusion**

In this work we investigated the effect of a variable bias anticathode on the electron kinetics in an electron beam generated E × B plasma. Measurements of the EEDFs indicate the electron population is nonthermal with beam component at the plasma center. In the plasma periphery, the beam component disappears and the EEDF is approximately a Maxwellian distribution. By varying the anticathode bias so that the anticathode is electron repelling, the axial losses of plasma to the anticathode are reduced, leading to increased plasma density.

The experimentally determined plasma density variation with the anticathode voltage bias was analyzed using a 0D electron continuity model. The anticathode bias voltage controls the anticathode sheath potential drop, which reflects electrons with insufficient energy to overcome the sheath potential. The model indicates that anomalous cross field transport is suppressed when the anticathode is electron collecting. This enhanced radial electron transport shorts the radial electric field, potentially reducing the ion confinement. In this way, the variable bias anticathode can be used to control axial and anomalous radial electron transport in the discharge

We have also demonstrated control of the EEDF in the e-beam E × B plasma system by varying the anticathode bias in the low-pressure operating regime of 0.1 mTorr. A warm electron population in the 10-30 eV range is present in the collector mode, while a bump on tail feature may be realized by operating in repeller mode. We propose that these EEDF features are the result of the beam-plasma system operating in either WT or SLT regimes, and that the anticathode bias may determine the turbulent regime by controlling the beam-plasma density ratio. In such a way, beam plasma instabilities lead to a redistribution of the EEDF, enhancing ionization in collector mode. Future studies should be conducted to further investigate the role of the anticathode voltage bias in controlling the instability regime of the plasma, potentially enabling selective producing EEDFs that are more suited for applications requiring enhanced ionization or radical production.

**Acknowledgements**

This research was supported by the U.S. Department of Energy, Office of Fusion Energy Science, under Contract No. DEAC02-09CH11466, as a part of the Princeton Collaborative Low Temperature Plasma Research Facility (PCRF). The authors are grateful to Andrei Smolyakov, Alexandre Likhanskii, Joseph Abbate, Sierra Jubin, and Sunghyun Son, and Igor Kaganovich for fruitful discussions regarding the physics of electron beam generated E×B plasmas and cross field transport and Timothy K. Bennett for technical support on the electron beam chamber.

**Appendix**

1. **Evidence of enhanced plasma oscillations in repeller mode**

The plasma density fluctuations in the plasma were determined using a negatively biased probe (ion probe) array inserted at $r = 1$ cm. Each probe in the probe array consisted of a tungsten wire with exposed length of 4.5 mm and diameter of 1.3 mm, biased with a battery pack to a fixed potential of $V_b = -100$ V relative to the grounded chamber. To avoid the effects of fluctuations in the electron temperature, the potential $V_b$ was chosen to be sufficiently negative such that the probe was collecting orbital motion limited (OML) ion current. This current collection regime was independently verified using a swept voltage on the probe to identify the ion saturation current collecting region of the probe current-voltage characteristic. Under such

OML conditions, and assuming quasineutrality, the ion current $I_i$ collected by the ion probe is given by

$$I_i(t) = \left(\frac{2}{\sqrt{\pi}} eA_{pr}\sqrt{-\frac{eV_b}{2\pi m_i}}\right) n_i(t), \tag{16}$$

where $A_{pr}$ is the probe exposed area [56]. Therefore, the ion current is proportional to the plasma density $n_i$. In order to quantify the plasma density fluctuation amplitude, we introduce the definition $\tilde{n}_i(t) = n_i(t) - n_{i0}$, where $n_{i0}$ is the time averaged plasma density and $\tilde{n}_i(t)$ is the fluctuating component of the plasma density, The effect of the anticathode bias on the plasma density fluctuation is shown in Figure 15. Evidently, the amplitude of the plasma density fluctuation is enhanced in the repeller mode over the collector mode. Additionally, comparing Figure 15a and b, the amplitude of the plasma density fluctuation is suppressed at the higher-pressure condition of $p = 0.5$ mTorr, indicating suppressed anomalous transport. The cases with larger plasma density fluctuations are associated with larger anomalous parameter $\alpha$ [11].

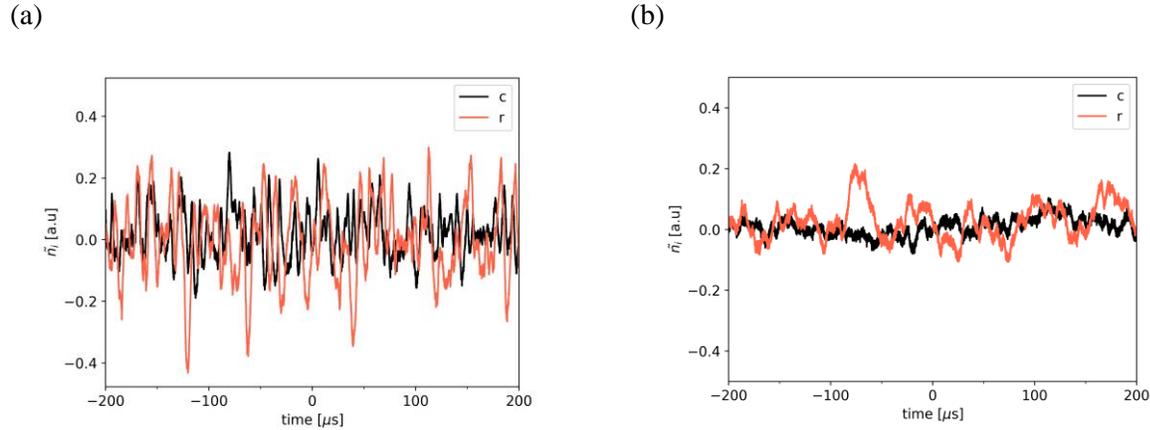

**Figure 15.** Measurement at $r = 1$ cm of oscillating component of plasma density vs. time for collector (c) and repeller (r) modes, for (a) $p = 0.1$ mTorr and (b) $p = 0.5$ mTorr. Other experimental parameters are held fixed at $B = 100$ G, and $V_c = -55$ V.

The suppression of plasma density fluctuations in the 0.5 mTorr regime qualitatively illustrates the effect of increased neutral pressure on the suppression of the azimuthally rotating spoke instability, which corresponds to a reduction in anomalous cross-field electron transport [7]. Indeed, previous works indicate that the enhancement of the cross-field transport by anomalous collisions tends to decrease as fluctuations and associated instabilities decrease in amplitude, which can occur at higher pressures where electron-neutral collisional damping of the instabilities increase in magnitude [8,16]. Under such conditions, the anomalous parameter $\alpha$ decreases as the pressure increases, leading to a reduction in the electron cross-field mobility (Eq. (11)). In this work, the axial current ratio increases as the pressure increases from $p = 0.1$ mTorr to $p = 0.5$ mTorr (Figure 3). Additionally, the experimentally determined plasma potential profiles outside of the ionization region, for $r \geq 2$ cm, all tend to decrease in value when increasing the neutral pressure from $p = 0.1$ mTorr to $p = 0.5$ mTorr (Figure 9). These trends both are consistent with reduced $\alpha$ and hence reduced radial anomalous cross-field electron transport at the higher-pressure condition.

### 2. Sign of plasma potential relative to anode

In the described experiments, the plasma potential is positive with respect to the anode (Figure 9). This is in contrast to the negative plasma potential observed in a similar $E \times B$ plasma systems [7]. This is likely due to the larger axial electron losses in the setup considered in this work. As described in Section II, the cathode region is restricted to $r < 1.75$ cm, which is significantly smaller than the cathode region of $r < 7.5$ cm in Ref. [7]. As a result, the area of plasma that is attached to the axial boundary on the cathode side of the chamber is larger in the present work, leading to a larger net loss of negative charge and hence a more positive plasma potential.

An additional corroborating measurement of $V_{pl}$ was performed independently using the LP, with the main probe biased to a potential $V_B$ using a 100V-2A Kepco BOP programmed with a 10Hz negative ramp voltage output. In this case, the voltage across a 10 k$\Omega$ shunt resistor placed in series with the LP was measured to determine the probe current $I_{pr}$. For each of these probe acquisitions, multiple current-voltage sweeps were obtained for statistical significance of data. The plasma potential was then determined by finding the maximum of $dI_{pr}(V_B)/dV_B$. The positive $V_{pl}$ with respect to the anode indicated by emissive probe measurements (Figure 9) are corroborated by measurements using the cold LP with sweeping voltage bias, which measured a plasma potential of $V_{pl} = 3 \pm 2$ V at $r = 0$ cm.

### 3. One-way flux of non-Maxwellian electrons, short-circuit effect, ion-induced SEE, and Ar metastable induced SEE

In this section we will discuss the electron flux into and out of the plasma control volume depicted in Figure 12 and discussed in Section V.1. We will estimate these fluxes using measured EEDF and explain why the assumption of Maxwellian EEDF for estimations of these fluxes (Section V.1) is justified. In addition, we will also show that electron fluxes due to short-circuit by equipotential plasma boundaries, ion-induced SEE, and metastable-induced SEE do not have a significant effect on the radial transport across the magnetic field.

The one-way axial flux of electrons can be approximated by the electron thermal flux, as given by Eq. (6). However, for significantly non-Maxwellian EEDFs, this flux might significantly deviation from the total one-way axial electron flux $\Gamma_{e,\text{tot.}}^{1-\text{way}}$. Here we will quantify how significant this deviation is.

Figure 16 shows the experimentally determined EEDF $f_{e,\text{expt.}}$ for parameters $p = 0.1$ mTorr, $B = 100$ G, $V_c = -55$ V, and $V_{atc} = -55$ V. A Maxwellian fit $f_{e,\text{fit}}^M$ is made to approximate the cold bulk electron population of $f_{e,\text{expt.}}$,

$$f_{e,\text{fit}}^M(\varepsilon) = \frac{2}{\sqrt{\pi}} \frac{n_{e,\text{fit}}}{\left(kT_{e,\text{fit}}\right)^{1.5}} \varepsilon^{1/2} \exp\left(-\frac{\varepsilon}{kT_{e,\text{fit}}}\right), \tag{17}$$

with fitting parameters $n_{e,\text{fit}} = 0.8 \times 10^{10}$ cm$^{-3}$ and $T_{e,\text{fit}} = 5$ eV. The difference between the experimentally measured EEDF and Maxwellian fit, $g = f_{e,\text{expt.}} - f_{e,\text{fit}}^M$, corresponds the non-Maxwellian component of $f_{e,\text{expt.}}$. Finally, an effective Maxwellian fit $f_{e,\text{eff.}}^M$ is shown, which is given by the same functional form as Eq. (17), but with electron density and temperature calculated directly from $f_{e,\text{expt.}}$ as described by Eqs. (2) and (3) (Figure 7 and Figure 8).

To compute the one-way flux associated with the non-Maxwellian component $g$, we make the simplifying assumption that the non-Maxwellian electrons only have velocity in the axial direction, such that the VDF associated with $g$ is one-dimensional. This assumption is justifiable under the consideration that the injected e-beam is magnetized perpendicular to the applied axial magnetic field. Furthermore, this assumption gives a conservative (maximum) estimate of the one-way electron flux in the axial direction. Under such an assumption, the one-way electron flux of the non-Maxwellian component $\Gamma_e^g$ is given by half of the total non-Maxwellian electron flux [43]

$$\Gamma_e^g = \frac{1}{2} \int_0^\infty \left(\frac{2\varepsilon}{m_e}\right)^{1/2} g(\varepsilon)\, d\varepsilon. \tag{18}$$

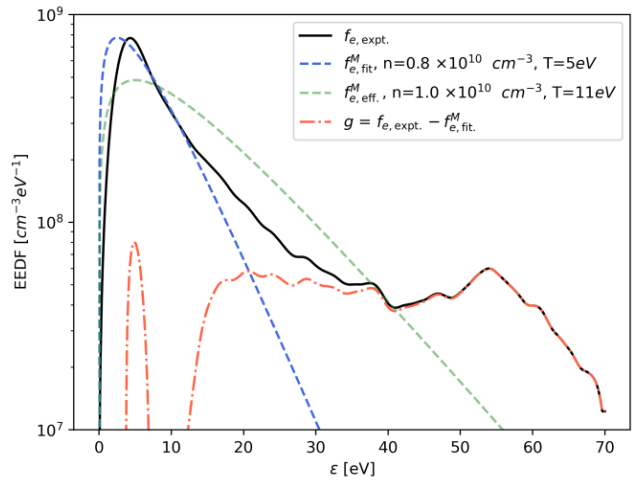

**Figure 16.** Experimentally determined EEDF $f_{e,\text{expt.}}$ at $r = 0$ cm for parameters $p = 0.1$ mTorr, $B = 100$ G, $V_c = -55$ V, and $V_{atc} = -55$ V (repeller mode). Also plotted is a Maxwellian fit $f_{e,\text{fit}}^M$ to the cold bulk electron population of $f_{e,\text{expt.}}$, the effective Maxwellian fit $f_{e,\text{eff.}}^M$ given by the measured effective electron density and temperature as described by Eqs. (2) and (3), and the difference between the experimentally measured EEDF and Maxwellian fit, $g = f_{e,\text{expt.}} - f_{e,\text{fit}}^M$.

The total one-way electron flux can be approximated as the sum of Maxwellian and non-Maxwellian one-way fluxes,

$$\Gamma_{e,\text{tot.}}^{1-\text{way}} = \Gamma_{e,\text{fit}}^M + \Gamma_e^g, \tag{19}$$

where the one-way flux associated with the Maxwellian fit is given by $\Gamma_{e,\text{fit}}^M = \frac{1}{4}n_{e,\text{fit}}(8kT_{e,\text{fit}}/\pi m_e)^{1/2}$. Table II shows each term of Eq. (19), as well as the one-way thermal flux approximated by the effective electron density and temperature given by Eq. (6), $\Gamma_{e,\text{eff.}}^M$. Despite the non-Maxwellian component $\Gamma_g$ contributing a significant portion of the total one-way flux, the effective thermal flux accurately predicts the total one-way flux within a factor of 1.2, such that $\Gamma_{e,\text{tot.}}^{1-\text{way}} \approx \Gamma_{e,\text{eff.}}^M$. Therefore, it is justifiable to use the effective one-way flux in the model in Section V.1 to approximate the one-way electron flux entering either the cathode or anticathode sheath.

**Table II.** The one-way electron flux and its components as given by Eqs. (18), and (19), as well as the effective one-way thermal flux given by Eq. (6).

| $\Gamma_{e,\text{tot.}}^{1-\text{way}}$ $[m^{-2}s^{-1}]$ | $\Gamma_{e,\text{fit}}^M$ $[m^{-2}s^{-1}]$ | $\Gamma_e^g$ $[m^{-2}s^{-1}]$ | $\Gamma_{e,\text{eff.}}^M$ $[m^{-2}s^{-1}]$ |
|---|---|---|---|
| $7 \times 10^{21}$ | $3 \times 10^{21}$ | $4 \times 10^{21}$ | $6 \times 10^{21}$ |

Another consideration regarding the confinement of electrons in the control volume is the so-called short circuit effect at the anticathode [44]. Because the anticathode and cathode surfaces are conducting electrostatic boundaries upon which the axial magnetic field lines terminate, electrons may effectively hop magnetic field lines by short-circuiting through the axial boundaries [44]. This short-circuit effect should manifest as an additional flux of electrons entering the cylindrical control volume considered in Figure 12, as short-circuiting electrons redistribute to cancel the ambipolar electric potential. However, the short circuit electron flux $\Gamma_e^{sc}$ entering a cylindrical end cap of the control volume is limited to be, at maximum, equal to the Bohm ion flux leaving the control volume to either of the axial electrode sheaths [57],

$$\Gamma_e^{sc,\text{max}} = n_e c_s \tag{20}$$

where $c_s = (kT_e/m_i)^{1/2}$ is the ion sound speed and we have assumed quasineutrality, $n_i \approx n_e$.

Similarly, the ion induced SEE flux emitted from the anticathode will also scale with the Bohm flux as

$$\Gamma_e^{iSEE,\text{max}} = \delta_i n_e c_s, \tag{21}$$

where $\delta_i$ is the ion induced SEE yield for Argon impinging stainless steel. For Ar ions incident on metallic surfaces with energy less than 100eV, $\delta_i \leq 0.1$ [41].

Metastable Ar has also been shown to have a significant SEE yield, with a yield $\delta_* \approx 1$ for Ar $1s_5$ metastables striking a stainless steel surface at room temperature [48]. The metastable induced SEE flux $\Gamma_e^{*SEE}$ evolved from the anticathode is

$$\Gamma_e^{*SEE} = \delta_* n_{1s_5} v_{th,n}, \tag{22}$$

where $n_{1s_5}$ is the Ar $1s_5$ density and $v_{th,n} = \left(\frac{8kT_n}{\pi m_{Ar}}\right)^{1/2}$ is the neutral Ar thermal velocity. Previous work on the same e-beam plasma considered in this work has shown that for identical operating conditions, in collector mode the neutral temperature at plasma center is close to room temperature, $T_n \approx 0.026$ eV [9]. Furthermore, in a similar e-beam generated plasma, the Ar $1s_5$ density was found to be roughly equal to the plasma density, with $n_{Ar1s_5} \approx 10^{16} m^{-3}$ [3]. In lieu of direct measurements of the metastable population in this work, we instead estimate $\Gamma_e^{*SEE}$ using this value of metastable density from Ref. [3].

Table III shows the relative contribution of each of the additional electron fluxes into the control volume compared to the volumetric ionization rate. The short-circuit, ion induced SEE, and metastable induced SEE fluxes are negligible compared to the measured total ionization rate in the control volume. Therefore, it is justifiable to omit these fluxes from the model described in Section V.1.

**Table III.** Electron fluxes into a cylindrical control volume of 1cm radius due to short-circuit effect, ion-induced SEE, and Ar metastable-induced SEE. Values are normalized to the measured volumetric ionization rate.

| $\dfrac{\Gamma_e^{sc,\max}}{R_{iz}L_{ch}}$ | $\dfrac{\Gamma_e^{iSEE,\max}}{R_{iz}L_{ch}}$ | $\dfrac{\Gamma_e^{*SEE}}{R_{iz}L_{ch}}$ |
|---|---|---|
| ~0.1 | ~0.01 | ~0.01 |

### 4. Beam plasma instability regimes

Here, we follow Ref. [19] and describe the boundaries of the beam-plasma instability regimes, demarcated by the transition between the weak turbulent (WT) and Strong Langmuir Turbulence (SLT) regimes. The results are shown in Fig. 14.

The boundary of the WT regime dominated by parametric decay instability (PDI) onset is given by

$$\frac{9}{8}\left(\frac{n_b}{n_e}\right)^{\frac{4}{3}}\frac{\varepsilon_b}{T_e} > 4\frac{\Delta_e}{\omega_{pe}}\frac{\Delta_i}{\omega_{IAW}}, \qquad (23)$$

and the boundary of the SLT regime dominated by standing wave modulational instability (SWMI) onset is given by

$$\frac{9}{8}\left(\frac{n_b}{n_e}\right)^{\frac{4}{3}}\frac{\varepsilon_b}{T_e} > \max\left[\frac{2\Delta_e}{\omega_{pe}}, (k\lambda_{De})^2\right]. \qquad (24)$$

Here, the electron and ion damping rates $\Delta_e$ and $\Delta_i$ are given approximately as [19,58]

$$\Delta_e \approx \sqrt{\frac{\pi}{8}}\frac{\omega_{pe}}{(k\lambda_{De})^3}\exp\left[-\frac{3}{2} - \frac{1}{2(k\lambda_{De})^2}\right] + \frac{\nu_{en}}{2}, \qquad (25)$$

$$\Delta_i \approx \sqrt{\frac{\pi}{8}} \frac{\omega_{IAW}}{(1+k^2\lambda_{De}^2)^{\frac{3}{2}}} \left\{ \sqrt{\frac{m_e}{m_i}} + \left(\frac{T_e}{T_i}\right)^{\frac{3}{2}} \exp\left[-\frac{3}{2} - \frac{\frac{T_e}{2T_i}}{(1+k_{IAW}^2\lambda_{De}^2)}\right] \right\} + \frac{\nu_{in}}{2}, \quad (26)$$

$$\omega_{IAW}^2 \approx \frac{k_{IAW}^2 c_s^2}{(1+k_{IAW}^2\lambda_{De}^2)}, \quad (27)$$

where $\omega_{IAW}$ and $k_{IAW}$ are the ion acoustic wave (IAW) frequency and wavenumber respectively, $c_s = (kT_e/m_i)^{0.5}$ is the Argon ion sound speed, $\lambda_{De}$ is the electron Debye length, and $\nu_{en}, \nu_{in}$ are the collision rates of electrons and ions with neutral Argon, respectively. The damping rate for electrons $\Delta_e$ is a combination of electrostatic wave Landau damping on electrons and the electron beam collisional damping with neutrals. The damping rate for ions $\Delta_i$ is similarly a combination of ion acoustic waves Landau damping on ions and the ions colliding with neutrals. Furthermore, the assumption is made that the IAW wavenumber is roughly condensed around the pump wave number, $k_{IAW} \sim k$.